\documentclass{emulateapj_latest}
\usepackage{apjfonts}

\journalinfo{The Astrophysical Journal, in press}
\slugcomment{Received 2007 December 17; accepted 2008 January 31}

\shortauthors{CAMILO ET AL.}
\shorttitle{RADIO-EMITTING MAGNETAR 1E~1547.0--5408}

\begin{document}


\def\mag{1E~1547.0--5408}
\def\psr{PSR~J1550--5418}
\def\snr{G327.24--0.13}
\def\xte{XTE~J1810--197}

\title{The magnetar 1E~1547.0--5408: radio spectrum, polarimetry, and timing}

\author{F.~Camilo,\altaffilmark{1}
  J.~Reynolds,\altaffilmark{2}
  S.~Johnston,\altaffilmark{3}
  J.~P.~Halpern,\altaffilmark{1}
  and S.~M.~Ransom\altaffilmark{4}
}

\altaffiltext{1}{Columbia Astrophysics Laboratory, Columbia University,
  New York, NY 10027.}
\altaffiltext{2}{Australia Telescope National Facility, CSIRO, Parkes
  Observatory, Parkes, NSW 2870, Australia.}
\altaffiltext{3}{Australia Telescope National Facility, CSIRO, Epping, 
  NSW 1710, Australia.}
\altaffiltext{4}{National Radio Astronomy Observatory, Charlottesville, 
  VA 22903.}

\begin{abstract}
We have investigated the radio emission from the anomalous X-ray pulsar
\mag\ (\psr) using the Parkes telescope and the Australia Telescope
Compact Array.  The flux density of the pulsar is roughly the same between
1.4 and 45\,GHz, but shows time variability.  The radiation is nearly
100\% linearly polarized between frequencies of 45 and 3.2\,GHz, but from
2.3 to 1.4\,GHz it becomes increasingly more depolarized.  The rotation
measure of $-1860$\,rad\,m$^{-2}$ is the largest for any known pulsar,
and implies an average magnetic field strength along the line of sight
of $2.7\,\mu$G.  The pulse profiles are circularly polarized at all
frequencies observed, more so at lower frequencies, at the $\approx 15\%$
level.  The observed swing of the position angle of linear polarization as
a function of pulse phase suggests that in this neutron star the rotation
and magnetic axes are nearly aligned, and that its radio emission is only
detectable within a small solid angle.  Timing measurements indicate that
the period derivative of this 2\,s pulsar has increased by nearly 40\%
in a 6-month period.  The flat spectrum and variability in flux density
and pulse profiles are reminiscent of the properties of \xte, the only
other known radio-emitting magnetar, and are anomalous by comparison
with those of ordinary radio pulsars.

\end{abstract}

\keywords{ISM: individual (G327.24--0.13) --- pulsars: individual
(1E~1547.0--5408, PSR~J1550--5418, XTE~J1810--197) --- stars: neutron}

\section{Introduction} \label{sec:intro} 

Anomalous X-ray pulsars (AXPs) and soft gamma-ray repeaters
are neutron stars whose long rotation periods (2--12\,s) are
a result of very strong surface magnetic fields (see Woods \&
Thompson 2006 for a review)\nocite{wt06}.  In the magnetar model
\citep[][1996]{dt92a,td95}\nocite{td96a}, their large and variable X-ray
luminosity is a result of magnetic field decay, but much remains unknown
about this class of young neutron stars.

Two of the 13 confirmed magnetars are radio emitters. The 5\,s AXP \xte\
shows remarkable pulsations \citep{crh+06} that have a flat spectrum
\citep{crp+07} and are highly linearly polarized \citep{crj+07,ksj+07}.
Its flux density and pulse profile vary greatly on short timescales
\citep{ccr+07}.  This radio emission \citep{hgb+05} arose only after an
X-ray outburst \citep{ims+04}.

The AXP \mag, recently identified at the center of the candidate supernova
remnant \snr\ \citep{gg07}, emits radio waves at the rotation period
$P=2$\,s \citep{crhr07}.  These are transient, and are apparently related
to the X-ray variability shown by the AXP \citep{hgr+07}.  Here we report
on a detailed study of the radio emission from \mag\ (\psr), and compare
its spectral, polarimetric, and timing properties to those of \xte.

\section{Observations, Analysis, and Results} \label{sec:obs}

\subsection{Radio Spectrum} \label{sec:flux}

\begin{deluxetable}{llrc}
\tablewidth{0.88\linewidth}
\tablecaption{\label{tab:atca} ATCA observations of \psr\ }
\tablecolumns{4}
\tablehead{
\colhead{Date}                 &
\colhead{Flux calibrator}      &
\colhead{Frequency}            &
\colhead{$S_\nu$}              \\
\colhead{}                     &
\colhead{}                     &
\colhead{$\nu$ (GHz)}          &
\colhead{(mJy)}
}
\startdata
2007 Jun 26\tablenotemark{a} & 0823--500 &  1.384 & $3.3\pm0.3$ \\
                             &           &  2.368 & $5.3\pm0.3$ \\
2007 Jul 10                  & 0823--500 &  1.384 & $4.4\pm0.4$ \\
                             &           &  2.368 & $5.4\pm0.3$ \\
                             &           &  4.800 & $6.4\pm0.3$ \\
                             &           &  8.640 & $5.5\pm0.2$ \\
2007 Jul 24                  & 1934--638 &  4.800 & $5.0\pm0.2$ \\
                             &           &  8.640 & $4.3\pm0.2$ \\
                             &           & 18.496 & $2.8\pm0.2$ \\
                             &           & 18.624 & $2.8\pm0.2$ \\
2007 Aug 20 & 1921--293\tablenotemark{b} & 42.944 & $5.6\pm1.2$ \\
                             &           & 44.992 & $5.6\pm0.7$
\enddata
\tablecomments{The array configuration was H168 for the last epoch,
and 6C for all others.  We used 1613--586 as the phase calibrator in all
instances.  In every case, at each frequency we sampled a bandwidth of
128\,MHz in each of two orthogonal linear polarizations, and visibilities
were accumulated into 32 pulsar phase bins. }
\tablenotetext{a}{Antennas CA01 and CA05 were off line. } 
\tablenotetext{b}{Uranus was observed following our session to tie the
flux scale to an absolute standard. }
\end{deluxetable}

We observed \psr\ on four occasions with the Australia Telescope Compact
Array (ATCA) interferometer, with the primary goal of measuring its
radio spectrum.  The observations were always done in pulsar-gated
mode using the known ephemeris, and in each case we obtained pulse
profiles with 32 phase bins in all Stokes parameters.  The first
observation, at simultaneous frequencies of 1.4 and 2.4\,GHz, was
already reported in \citet{crhr07}.  Details of all observations are
given in Table~\ref{tab:atca}.  Typically we first observed the flux
calibrator and then interleaved $\approx 1$--2 minute observations of
the phase calibrator with $\approx 5$--20 minute observations of the
pulsar (at higher radio frequencies we observed the phase calibrator
more frequently).  Two frequency bands were always observed simultaneously
(in the middle two epochs we interleaved observations at 4.8 and 8.6\,GHz
with the other frequency pair; see Table~\ref{tab:atca}).  The integration
times on the pulsar varied substantially, ranging between 2 and 6\,hr
for each frequency pair.  With the array configurations used, we were
in effect sensitive only to point sources.

\begin{figure}[t]
\begin{center}
\includegraphics[angle=0,scale=0.42]{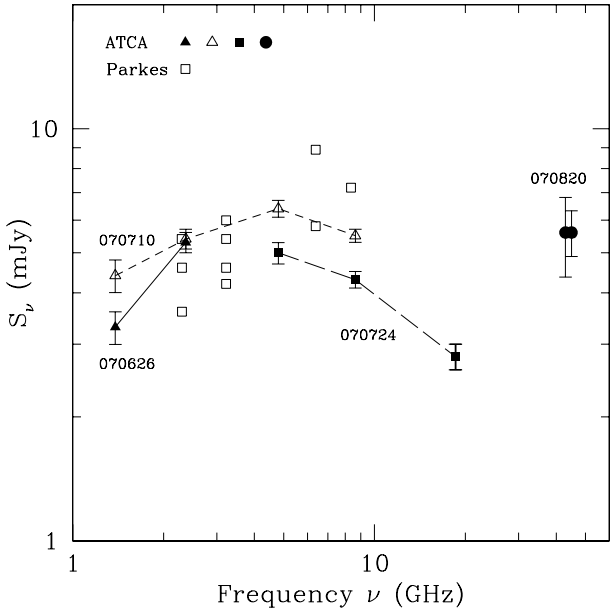}
\caption{\label{fig:flux}
Flux densities of \psr\ at 1.4--45\,GHz.  All points with error bars were
derived from data obtained at the ATCA on four separate dates, labeled.
Data collected at multiple frequencies on the same day are connected by
lines.  Open squares represent flux-calibrated measurements obtained at
Parkes with a digital filterbank (see text) and have typical fractional
uncertainties of about 25\%.  }
\end{center}
\end{figure}

We analyzed all data sets in a uniform manner
using standard techniques with MIRIAD\footnote{See
http://www.atnf.csiro.au/computing/software/miriad/.}.  The 7\,mm
(30--50\,GHz) system is new at the ATCA, and we took particular care
to obtain the reported fluxes.  The flux densities we used for the flux
calibrator, 10.65\,Jy at 43\,GHz and 10.72\,Jy at 45\,GHz, were obtained
immediately following our observations in a monitoring program using
Uranus to tie 7\,mm fluxes to an absolute scale.  We also used the option
``opcorr'' in the MIRIAD task {\tt atlod} in order to determine fluxes
corrected for atmospheric opacity.  All ATCA flux density measurements
and uncertainties (Table~\ref{tab:atca} and Fig.~\ref{fig:flux}) were
obtained with the task {\tt imfit}.

Using the Parkes radio telescope we observed \psr\ with a digital
filterbank \citep[PDFB1 or PDFB2;][]{man06} at five different frequencies,
on a total of 12 occasions.  The main goal of these observations was to
obtain calibrated polarimetric profiles (\S~\ref{sec:pol}), but we also
measured the flux densities.  The Hydra~A flux standard was observed
at each epoch, and the data were analyzed with PSRCHIVE \citep{hvm04}.
The resulting flux density measurements have an estimated fractional
uncertainty of $\sim 25\%$, dominated by systematics in profile baselines,
and are shown in Figure~\ref{fig:flux}.

It is clear from Figure~\ref{fig:flux} that the flux density of \psr\
varies with time.  With an integrated column density of free electrons
$\mbox{DM}=830$\,cm$^{-3}$\,pc, this variability cannot be ascribed
to interstellar scintillation but rather is intrinsic \citep{crhr07}.
At each frequency for which multiple observations exist, the variability
appears to be of order $\approx 30\%$ about the mean, but can reach
$\sim 50\%$.

Most ordinary pulsars have steep radio power law spectra \citep{lylg95},
with $\alpha < -1$ ($S_{\nu} \propto \nu^{\alpha}$).  The spectrum
of \psr\ is very different.  It is remarkably flat over the range
1.4--45\,GHz, but apparently cannot be described by one spectral index.
Neglecting the data points at $\nu \approx 44$\,GHz, the spectrum could
be described as approximately log parabolic with a peak at $\approx
6$\,GHz.  However, the precise description of the spectrum at $\nu >
9$\,GHz requires greater reliability than is provided by the single-epoch
measurements at $\approx 18$ and 44\,GHz, and further observations are
required to clarify this.

\subsection{Polarimetry} \label{sec:pol}

\begin{figure}[t]
\begin{center}
\includegraphics[angle=0,scale=1.04]{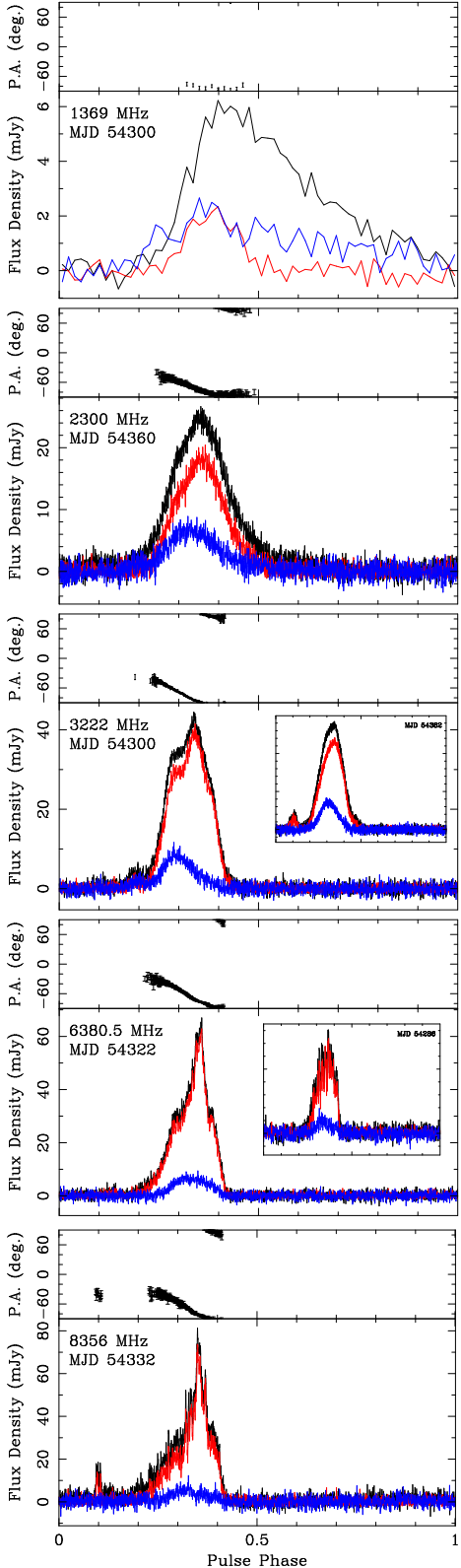}
\caption{\label{fig:pol}
Representative polarimetric profiles of \psr\ at frequencies of
1.4--8.4\,GHz.  The black, red, and blue lines represent, respectively,
the total intensity and the amount of linear and circular polarization
(Stokes parameters $I$, $L=(Q^2+U^2)^{1/2}$, and $V$).  The position
angles ($\mbox{P.A.} = \frac{1}{2} \arctan\ (U/Q)$) are shown for bins
where the $L$ signal-to-noise ratio $>4$, and have been corrected for
RM to their values at infinite frequency.  The 1.4\,GHz profiles are
displayed with 64 phase bins; all others, with 2048 bins.  }
\end{center}
\end{figure}

We have obtained polarimetric data from \psr\ at Parkes on 12 occasions
between 2007 July 5 and October 9.  We used five different feed/receiver
combinations: central beam of the 20\,cm multibeam receiver (1.4\,GHz),
Galileo (2.3\,GHz), 10/50\,cm (3\,GHz), central beam of the methanol
multibeam receiver (6.4\,GHz), and Mars (8.4\,GHz).  For the most recent
3\,GHz observation (inset in Fig.~\ref{fig:pol}) we recorded data with a
1\,GHz bandwidth digital filterbank (PDFB2); at 8.4\,GHz we used PDFB2
with 512\,MHz of bandwidth.  In all other instances we recorded 256\,MHz
of band with PDFB1.  Typically $\sim 10$ pulse periods were folded in
each subscan with 2048 bins across the pulse profile (1\,ms resolution),
and scans lasting up to $\sim 1$\,hr were interspersed with $\sim 1$
minute observations of a pulsed calibrating signal in order to determine
the relative gains and phases of the two feed probes in each receiver.
At each epoch we also observed the Vela pulsar in order to provide
a check on our polarimetric calibration, and Hydra~A in order to
obtain calibrated flux density measurements (see \S~\ref{sec:flux}).
All data were analyzed with PSRCHIVE.  Further details concerning the
instrumentation and data analysis can be found in \citet{crj+07}.

Faraday rotation affects polarized radiation as it traverses the
magnetized ISM.  We determined the rotation measure of \psr\ by
measuring its position angle of linear polarization (P.A.) as a function
of frequency within each 256\,MHz band at 2.3, 3.2, and 6.4\,GHz.
The resulting $\mbox{RM} = -1860 \pm 20$\,rad\,m$^{-2}$ is the largest
known for any pulsar.  Together with the DM, it implies an average
magnetic field strength along the line of sight weighted by electron
density of $1.2\,\mbox{RM}/\mbox{DM} = 2.7\,\mu$G.  Near this line of
sight, $(l,b) = (327\arcdeg,0\arcdeg)$, a number of compact radio sources
have $\mbox{RM} \approx -1000$\,rad\,m$^{-2}$ \citep{gdm+01}, so that
the large-scale Galactic field in this direction may be able to account
for the extreme RM of \psr.  Also near this direction, all pulsars at
distance $\la 5$\,kpc have very small RM \citep{hml+06}, suggesting that
\psr\ is more distant, compatible with the DM-derived $d \approx 9$\,kpc.

In Figure~\ref{fig:pol} we show representative polarimetric profiles of
\psr\ at 1.4, 2.3, 3.2, 6.4, and 8.4\,GHz (as shown by the insets at 3.1
and 6.4\,GHz, sometimes the profiles vary).  In each, the RM has been
used to correct the frequency-integrated linear polarization and the
observed P.A. to infinite frequency.  It is evident from the figure that
usually at $\nu \ge 3$\,GHz the emission is $>90\%$ linearly polarized
throughout the profile.  At lower frequencies, however, the fraction
of linear polarization is reduced: to about 80\% at 2.3\,GHz, and 25\%
at 1.4\,GHz.  At these frequencies the profile is scatter-broadened
due to mutipath propagation in the ISM \citep{crhr07}, and these two
facts are partly related (see \S~\ref{sec:depol}).  The profiles also
show substantial circular polarization, the level of which decreases
with increasing frequency, from $\ga 30\%$ at 1.4\,GHz to $\la 10\%$
at 8.4\,GHz.

The ATCA observations (\S~\ref{sec:flux}) extend these results to
higher frequencies: at $\nu \approx 18$ and 44\,GHz, the radiation
from \psr\ remains approximately 100\% linearly polarized throughout
the profile (which, at $\approx 2$ bins FWHM out of 32, appears to be
somewhat narrower than at lower frequencies), and the level of circular
polarization is $\la 10\%$.

\begin{figure}[t]
\begin{center}
\includegraphics[angle=270,scale=0.35]{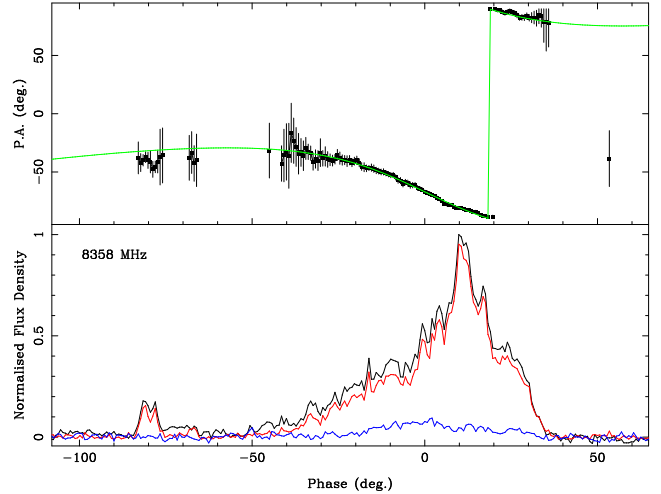}
\caption{\label{fig:rvm}
{\em Bottom}:  Polarimetric profile of \psr\ at 8.4\,GHz with 170 degrees
of pulse phase shown.  The black trace shows the total intensity, with the
red and blue showing linear and circular polarization respectively.  {\em
Top}:  P.A. data (black points) and the RVM assuming $\alpha=160^\circ$,
$\beta=14^\circ$.  Here we are using the definitions of $\alpha$
and $\beta$ given by \citet{ew01}.  Here and in Fig.~\ref{fig:multi},
the profiles are displayed with 512 bins per pulse period, and P.A. is
plotted when the linear polarization signal-to-noise ratio $>2.5$.
}
\end{center}
\end{figure}

If certain assumptions are met (principally a dipole field geometry),
the swing of P.A. as a function of pulse phase allows us to determine
the geometry of the star, in particular the angle between its rotation
and magnetic axes ($\alpha$) and the impact parameter of the line of
sight to the magnetic pole ($\beta$).  Figure~\ref{fig:rvm} shows the
results of fitting the P.A. curve at 8.4\,GHz to the rotating vector model
\citep[RVM;][]{rc69a}.  Because of the wide longitude range over which
this pulsar emits, constraints from the RVM fitting are good.  We find
that $\alpha$ must be greater than $140^\circ$ (i.e., that the pulsar
is almost aligned) and that $\alpha+\beta$ must be close to $180^\circ$,
so that the line of sight remains largely within the beam at all times.
This is confirmed by obtaining very similar results using data at 3\,GHz
that also show the presence of the initial leading component.

\begin{figure}[t]
\begin{center}
\includegraphics[angle=270,scale=0.35]{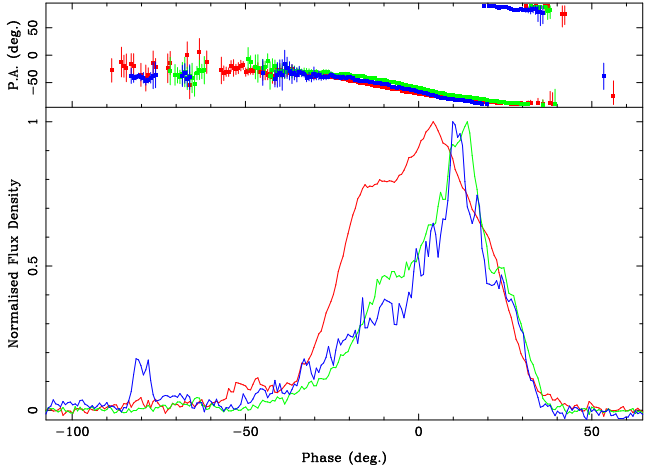}
\caption{\label{fig:multi}
Total intensity profiles ({\em bottom}) and P.A.s ({\em top}) for
\psr\ at three frequencies (red corresponds to 3.2\,GHz data, green to
6.4\,GHz, and blue to 8.4\,GHz).  The P.A.s are corrected for an RM of
--1860\,rad\,m$^{-2}$ and represent the (frequency independent) values at
the pulsar.  The point of zero phase is located at the steepest gradient
of the P.A. swing.
}
\end{center}
\end{figure}

Figure~\ref{fig:multi} shows the alignment of the profiles at 3.2, 6.4
and 8.4\,GHz.  Zero phase has been assigned from the RVM fit at 3.2 and
8.4\,GHz, with the 6.4\,GHz profile aligned by eye.  A number of features
are readily apparent.  The bulk of the emission covers some $80^\circ$
of pulse phase, with the midpoint at the location of the steepest swing of
P.A..  At 3.2\,GHz there are at least three main pulse components visible:
one near $-20^\circ$, one near $+5^\circ$, and one near $+25^\circ$, with
each of the three having roughly equal amplitude.  At 6.4 and 8.4\,GHz,
the first component is substantially reduced in amplitude compared to the
other two.  Finally, at 8.4\,GHz and occasionally at 3\,GHz a precursor
component is seen at phase $-80^\circ$.  The alignment of the position
angles at all three frequencies is remarkable especially over such a
large longitude range.

\subsubsection{Depolarization} \label{sec:depol}

The pulses from \psr\ are close to 100\% linearly polarized at $\nu \geq
3$\,GHz, with a modest component of circular polarization that decreases
with increasing frequency.  The linear polarization is evidently smaller
at 2.3\,GHz, and strikingly more so at 1.4\,GHz (Fig.~\ref{fig:pol}).
While this frequency dependence of polarization may be an intrinsic
property of the pulsar, we must also consider the effects of interstellar
scattering, which severely broadens the pulse at 1.4\,GHz with a
1\,s timescale \citep[see top panel of Fig.~\ref{fig:pol};][]{crhr07}.
At 2.3\,GHz, the pulse appears to be only slightly broadened in comparison
with higher frequencies, which is consistent with the expected $\nu^{-4}$
dependence of the scattering timescale for scatterers of a single
characteristic size: $\tau_{1.4} \approx 1$\,s at 1.4\,GHz corresponds
to $\tau_{2.3} \approx 0.14$\,s at 2.3\,GHz.

\begin{figure}[t]
\begin{center}
\includegraphics[angle=0,scale=0.43]{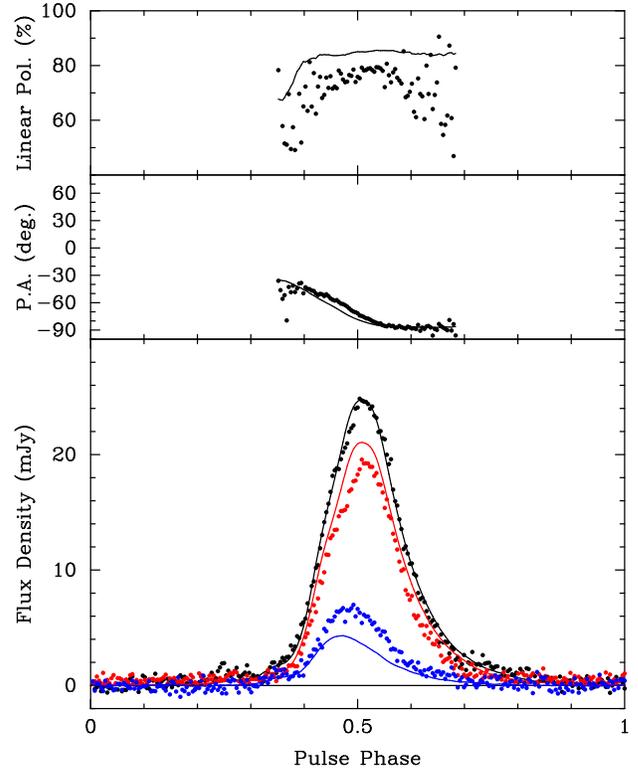}
\caption{\label{fig:scatt}
Observed 2.3\,GHz profiles of \psr\ ({\em dots}) and a model that consists
of the 3.2\,GHz Stokes parameters convolved with an exponential of
$\tau_{2.3} = 0.145$\,s ({\em solid lines}).  The parameter $\tau_{2.3}$
was adjusted by eye to give a good fit to total intensity.  Colors in
the bottom panel are as in Fig.~\ref{fig:pol}, and the relevant 3.2 and
2.3\,GHz profiles are also shown in that figure.  P.A. is plotted here
between $-100^{\circ}$ and $+80^{\circ}$ in order for the data not to
wrap around as it does in Fig.~\ref{fig:pol}.
}
\end{center}
\end{figure}

Scattering is expected to reduce the linear polarization because the
P.A. rotates through the different phases of the pulse, which are blended
at the observer.  This effect can be modeled at 2.3\,GHz by assuming that
the intrinsic pulse is similar to that observed at 3.2\,GHz.  Following
\citet{lh03} we convolve the individual Stokes parameters $I,Q,U,V$ at
3.2\,GHz with the function $g(t)= \exp(-t/\tau_{2.3})/\tau_{2.3}$ for $t >
0$, representing a thin scattering screen, varying $\tau_{2.3}$ to best
fit the total intensity profile at 2.3\,GHz.  The result of this model is
shown in Figure~\ref{fig:scatt}.  A good representation of the 2.3\,GHz
$I$ profile is obtained with $\tau_{2.3} = 0.145$\,s.  The expected
effects on the polarization are clearly seen in the model, namely, a
reduction in the percentage of linear polarization, $(Q^2+U^2)^{1/2}/I$,
and a flattening of the P.A. curve on the trailing side of the pulse.
But scattering does not fully account for the observed reduction in
linear polarization at 2.3\,GHz.  The model predicts maximum linear
polarization of 85\% at the place where the observed is 79\% (see top
panel of Fig.~\ref{fig:scatt}).  The data also show a flattening of
the P.A. angle on the trailing side of the pulse, although it does not
quite match the model on the leading side.  We also applied a model
of scattering from a medium extended between the source and observer
\citep{wil72,lh03}, which produces similar results and therefore is not
shown here.  At 1.4\,GHz the difference between similar scattering models
and the observed profiles is much greater.  This is easy to understand:
if the range of P.A.s emitted at 1.4\,GHz is as limited as those observed
at $\nu \ge 3$\,GHz, no amount of scattering can mix those P.A.s to
yield the very low level of linear polarization observed at 1.4\,GHz
(top panel of Fig.~\ref{fig:pol}).

Another interstellar effect that may reduce the linear polarization,
without affecting the circular component, is deviations in RM over the
different paths taken by the scattered rays.  Non-uniformity of either the
electron density or the magnetic field on transverse length scales $a$
corresponding to the scattering timescale, $a \approx (cd\tau)^{1/2}
= 1$--$3 \times 10^{16}$\,cm, if large enough to change RM by $\sim
60$\,rad\,m$^{-2}$, or 3\% of its measured value, would severely reduce
linear polarization at 1.4--2.3\,GHz by mixing the P.A.s.  However,
we do not explicitly model this ad hoc scenario here.

In summary, a scattering model accounts for only part of the observed
reduction in linear polarization at low frequencies, and it has little
effect on circular polarization.  There is also an intrinsic effect
in which circular polarization decreases with increasing frequency
while linear polarization increases.  Among ordinary pulsars, circular
polarization increases with increasing frequency for some, but in others
it decreases \citep{yh06}.


\subsection{Timing} \label{sec:timing}

\begin{figure}[t]
\begin{center}
\includegraphics[angle=0,scale=0.43]{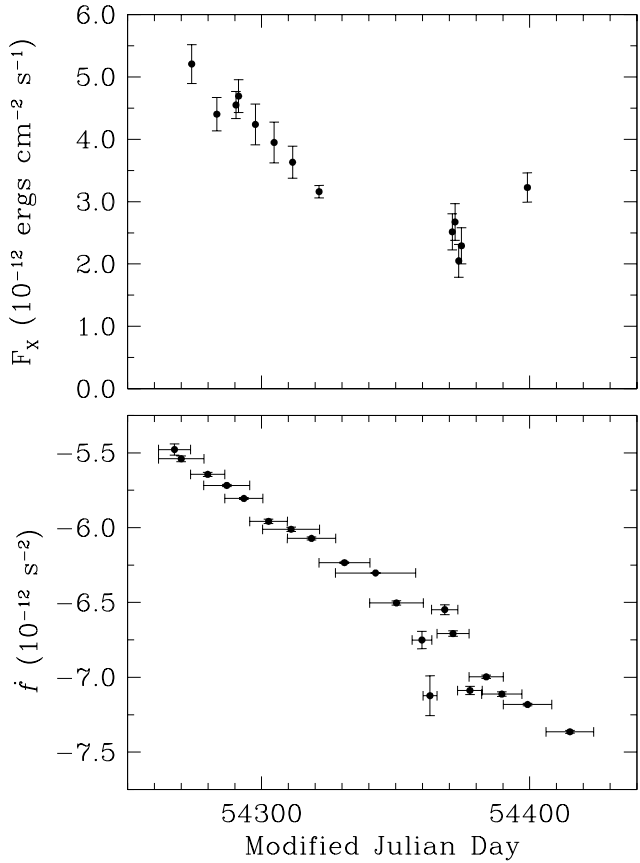}
\caption{\label{fig:timing}
{\em Bottom}: Evolution of spin frequency derivative for \psr.  All the
data were obtained at Parkes, and each point is from a TEMPO fit to
$f$ and $\dot f$.  After decreasing at a constant rate for 3 months,
between MJD 54360 and 54365 $\dot f$ decreased much more rapidly, and
after fluctuating over the following 2 weeks, it has continued its trend.
{\em Top}: Absorbed, 1--8\,keV X-ray flux of \mag\ \citep[from][]{hgr+07}.
}
\end{center}
\end{figure}

Since the discovery of pulsations from \mag\ on 2007 June 8, we have
timed the pulsar at Parkes on a regular basis.  As of November 22, we
have obtained times-of-arrival (TOAs) on 36 separate days.  These have
been derived from data collected with the analog filterbank used to
discover pulsations, at all frequencies mentioned in \S~\ref{sec:pol}
according to receiver availability.  Although the pulse profiles are
different at separate frequencies (and sometimes vary even at one
frequency; see Fig.~\ref{fig:pol}), this does not affect greatly the
overall timing precision, which is dominated by rotational instabilities
in the neutron star.

We have tracked the rotational evolution of \psr\ using TEMPO\footnote{See
http://www.atnf.csiro.au/research/pulsar/tempo/.} to fit the TOAs to
models that include rotational phase, spin frequency $f=1/P$, and $\dot f$
\citep[the celestial coordinates were held fixed at the values obtained
from ATCA data; see][]{crhr07}.  When the data span exceeds approximately
1 month, these models no longer describe well the behavior of the star.
As is common for magnetars that display substantial ``timing noise,''
the addition of higher frequency derivatives to the model can improve
the timing fit, although without predictive value.  Instead here, in
Figure~\ref{fig:timing}, we present the results of piece-wise fits to $f$
and $\dot f$.  In order to better sample the variation of $\dot f$, the
time intervals used to obtain these fits were made as short as possible
consistent with a relatively small uncertainty for $\dot f$.

For the first 3 months of our monitoring, $\dot f$ decreased at a constant
rate ($\ddot f = -1.21\times10^{-19}$\,s$^{-3}$), for a total decrease
of about 20\% compared to the initial value.  Then, sometime within
MJDs 54360--54365, $\dot f$ quickly decreased further, and within the
subsequent 2 weeks fluctuated before continuing its decreasing trend at
the same rate as before (see bottom panel of Fig.~\ref{fig:timing}).
There is no indication that any permanent period step resulted from
the interval of fluctuating $\dot f$.  As of late 2007 November, the
characteristic age of \psr\ is $P/2 \dot P = 1$\,kyr.

\section{Discussion} \label{sec:disc} 

In the 6 months since the discovery of pulsations from \psr, its period
derivative and torque have increased by about 40\%.  In magnitude,
this level of ``timing noise'' is not unprecedented for magnetars.
Quantifying timing noise as the magnitude of its cumulative contribution
to the cubic term of a Taylor series expansion of rotational phase
over a time interval $t$, i.e., $\ddot f t^3/6$ \citep{antt94}, this
amounts to about 60 cycles over 6 months, compared to 20 cycles over
9 months for the 5.5\,s AXP \xte\ \citep{ccr+07}.  However, the torque
of \xte\ was decreasing, at a time when the neutron star was returning
to quiescence years after a large outburst.  As the torque decreased,
so did the radio flux, which could indicate a decrease of particle flux
in the magnetosphere that might itself be partly responsible for the
decreased torque \citep[see in this regard][]{klo+06}.

In \psr, the torque has been increasing, at a time when the X-ray flux
has been generally decreasing (see Fig.~\ref{fig:timing}).  It appears
that after the $\dot f$ fluctuated in late 2007 September, the X-ray flux
stabilized or even increased, but poor sampling of the light curve makes
it hard to establish with confidence the correlative rotational--radiative
behavior. 

In any case, it remains unclear why torque should continue to increase
while X-ray luminosity decreases.  A logical explanation for the large
excursions in spin-down rate of magnetars involves an extra torque
(in addition to the magnetic dipole spin-down) derived from a particle
wind that both ``combs out'' the magnetic field and carries additional
angular momentum \citep{hck99,tdw+00}.  The luminosity of such winds
is considered to be derived from either ongoing seismic activity, or
static magnetospheric currents \citep{tlk02}, neither of which should
correlate inversely with X-ray luminosity.  (The radio luminosity, which
is energetically inconsequential but may point to a substantial particle
wind, so far shows no secular trend.)  Even more dramatic deviations
of torque evolution from X-ray luminosity were observed from the AXP
1E~1048.1--5937 \citep{gk04}: its X-ray flux increased by a factor of
4 over a period of 1 month, then decreased over 6 months to half its
peak value, and stayed relatively constant for 1 year thereafter; in
the meantime, the torque increased by a factor of 10 during the decay
of the X-ray flux, and subsequently varied by factors of up to 3 in
both directions during the period of stable X-ray flux.  We have been
observing \mag\ since 2007 June, at least 2 months after its radio flux
increased and up to 10 months after its X-ray flux increased by a factor
of $\ge 16$ \citep{hgr+07}, and it remains to be seen how its long-term
behavior compares to that of 1E~1048.1--5937.

Flux density variations such as observed in \psr, by up to $\sim
50\%$ on timescales of a few days (see Fig.~\ref{fig:flux}), which
cannot be accounted for by diffractive or refractive scintillation,
are intrinsic and are not observed in ordinary pulsars.  Also, the
variability observed in its pulse profiles, two examples of which are
shown in Figure~\ref{fig:pol}, is unlike that of ordinary pulsars.
Such changes are however observed in the other known radio magnetar,
\xte, where they are even more striking \citep{ccr+07}.  Although the
cause of these variations is not known, they may be related to changes
in magnetospheric plasma densities and/or currents that might also affect
the torque on the neutron star.

The radio spectrum of \psr\ (Fig.~\ref{fig:flux}) is remarkable (we
neglect here the apparent dip at $\nu \approx 18$\,GHz which, if real,
would make it even more remarkable; due to observed variability at lower
frequencies and only one observation at each of $\approx 18$ and 44\,GHz,
we consider this dip as tentative).  Although its flux density at 1.4\,GHz
is not particularly large (10\% of known pulsars have a larger value),
at 45\,GHz no pulsar is presently brighter \citep[see, e.g.,][]{kjdw97}.
In this regard, \psr\ is similar to \xte, which has an approximately flat
(and variable) spectrum between 0.3 and 144\,GHz \citep[$-0.5 \la \alpha
\la 0$;][]{crp+07}.  The reason for this is not known but it is also
not understood, after 40 years, why ordinary pulsars have steep spectra,
with $\langle \alpha \rangle = -1.6$.  The differences observed between
the radio spectra of ordinary pulsars and magnetars could be important
clues to both of their emission processes.  Whether they are caused,
e.g., by different particle population and acceleration mechanisms,
or by magnetospheric propagation effects, remains to be investigated.

The flat spectrum of \psr\ has allowed us to measure its polarimetric
properties up to 45\,GHz, a record among pulsars.  The slow sweep of
P.A. and its absolute value are identical at all frequencies observed
(see Fig.~\ref{fig:multi}), as expected in the RVM.  Also, the overall
pulse profile is wide and highly linearly polarized with relatively little
evolution with frequency (with the possible exception of low frequencies).
These characteristics are similar to those of the radio AXP \xte\
\citep{crj+07}, and to those of young ordinary pulsars \citep{jw06}.
Therefore, while unexpected a priori, it appears that ideas developed
to understand the geometry of emission from ordinary pulsars along open
dipolar magnetic field lines also apply to both known radio magnetars.

For \psr, the results of the RVM fits (\S~\ref{sec:pol}) imply that
the rotation and magnetic axes are nearly aligned.  Therefore, even
though the polar cap radius, $\propto P^{-1/2}$, is small, the observed
profile is wide because the line of sight remains largely within the
emission beam.  The radio beaming fraction for such a magnetar, however,
is expected to be small.  Conversely, for \xte\ we determined that the
most likely geometry is one with a substantial misalignment between
the rotation and magnetic axes, and a correspondingly large beaming
fraction \citep[and also emission height;][]{crj+07}.  For \xte, the
RVM fits also allow for a formally good solution with aligned axes, but
we prefer the misaligned geometry because it better explains the large
observed modulation of thermal X-rays \citep[this is shown in great
detail in the modeling of][]{pg08}.  For \psr, on the other hand, the
very small observed X-ray modulation \citep{hgr+07} is more compatible
with the nearly aligned geometry preferred from our RVM fits.

The aligned geometry does not constrain emission height.  However,
\citet{jw06} showed that emission from a cone high in the magnetosphere
could explain many observed properties of young pulsars, and the
properties of \psr\ are consistent with this picture.  Also unlike in
\xte\ \citep{crj+07}, it is not clear whether there is any evidence in
\psr\ that higher radio frequencies are emitted closer to the star than
lower frequencies, since the width of pulse components does not clearly
decrease with increasing frequency (see Fig.~\ref{fig:multi}).  A final
difference between the polarimetric characteristics of the two known
radio AXPs is that in \xte\ the level of circular polarization is very
small ($<10\%$) and does not change with frequency, while in \psr\ the
circular polarization is larger and increases with decreasing frequency,
as the linear polarization decreases.

Based on our study of the two known radio magnetars, we can identify
significant common properties:  both pulsars are very highly linearly
polarized; both have very flat spectra over a wide range of frequencies;
both have variable pulse profiles and radio flux densities.  The first
of these properties is also common with many ordinary young pulsars,
while the latter two are exceptional.   The cause of these traits is
not known, and their understanding would greatly increase our knowledge
of the magnetospheres of highly magnetized neutron stars.  It is also
not known why these are the only two magnetars so far to have been
detected at radio wavelengths.  \xte\ is clearly a transient AXP,
in the sense that in quiescence its surface temperature is as low as
that of some ordinary young neutron stars \citep{ghbb04}.  However,
this is not obviously the case with \mag\ \citep[see][]{hgr+07}, and
it may bear more of a resemblance to ``persistent'' AXPs.  Its radio
emission appears to make it unusual among AXPs, but it has a small
beaming fraction and is only transiently detectable \citep{crhr07}.
Therefore, without any existing theoretical basis for excluding radio
emission from magnetars, it seems reasonable to suppose that any magnetar
could in fact occasionally emit radio waves.

\acknowledgements
We thank Michael Dahlem for his generous help, particularly with
calibrating ATCA 7\,mm data, John Sarkissian for observing assistance,
and the dedicated staff at Parkes.  Willem van Straten kindly helped
with the PSRCHIVE software.  We are grateful to Phil Edwards and Dave
McConnell for quickly approving and scheduling the observations at
the ATCA.  The Parkes Observatory and the ATCA are part of the Australia
Telescope, which is funded by the Commonwealth of Australia for operation
as a National Facility managed by CSIRO.  This work was supported in
part by the NSF through grant AST-05-07376 to F.C.


\end{document}